\documentclass{article}
\usepackage{amsmath}

\renewcommand{\theequation}{\arabic{section}.\arabic{equation}}

\begin{document}

\title{Transverse-momentum dependent parton distributions in large-$N_{c}$
QCD}
\author{P.V.~Pobylitsa}
\date{}
\maketitle

\begin{center}
\emph{Institute for Theoretical Physics II, Ruhr University Bochum, D-44780,
Bochum, Germany\\[0pt]}
and\\[0pt]
\emph{Petersburg Nuclear Physics Institute, Gatchina, St. Petersburg, 188300,
Russia}
\end{center}

\begin{abstract}
The transverse-momentum dependent quark distributions are studied in the
limit of the large number of colors $N_{c}$. The leading orders of $N_{c}$
are determined for all twist-2 distributions including the $T$-odd
functions. The isospin structure of the dominant contributions is found.
\end{abstract}

\section{Introduction}

\setcounter{equation}{0}

Transverse-momentum dependent parton distributions appear in the theoretical
description of single spin asymmetries in various hard processes
including semi-inclusive deep inelastic scattering, Drell-Yan process and
inclusive pion production \cite{Sivers-90,ABM-95,BM-98,BM-99}. Although the
phenomenological applications of transverse-momentum dependent distributions
typically involve QCD-inspired methods, whose status does not reach the
standards of solid factorization theorems, certain aspects of this work have
been recently clarified by taking a more careful attitude to the
role played by the initial- and final-state interactions \cite%
{BHS-02,Collins-02,JY-02}. It has been emphasized \cite{Collins-02} that the
choice of the Wilson lines in the definition of the transverse-momentum
dependent parton distributions is sensitive to the structure of the initial-
and final-state interactions in the considered hard process. In particular,
this gives a theoretical argument in favor of nonvanishing $T$-odd
distributions.

The nonperturbative nature of the transverse-momentum dependent parton
distributions leaves us not too much space for a serious theoretical
analysis of their properties. One of the standard methods allowing to study
nonperturbative quantities in QCD is the limit of the large number of colors 
$N_{c}$ which was introduced in QCD by 't Hooft \cite{Hooft-74} and analyzed
for baryons by Witten \cite{Witten-79}. In this paper we study the large-$%
N_{c}$ limit of the twist-2 transverse-momentum dependent parton
distributions in nucleon, assuming the $SU(2)$ isospin symmetry. In particular,
we determine the dominant spin and isospin components of parton
distributions at large $N_{c}$. Certainly the $1/N_{c}$-expansion has its
limitations. The expansion parameter $1/N_{c}=1/3$ is not too small. Even in
the leading order of the $1/N_{c}$-expansion, QCD is not solved, so that
usually the work is reduced to the counting of powers of $N_{c}$ for various
quantities and to the identification of dominant and suppressed structures.
However, keeping in mind the general success of the phenomenological
applications of the $1/N_{c}$-expansion, we hope that our results obtained
for the transverse-momentum dependent distributions at $N_{c}\rightarrow%
\infty$ may have at least qualitative attitude to the real world.

The naive definition of the transverse-momentum dependent parton
distributions is based on the matrix element of the quark fields $\psi$ over
the nucleon state $|P,S\rangle$ with momentum $P$ and spin $S$%
\begin{equation}
P_{\mu}P^{\mu}=M^{2}\,,\quad S_{\mu}S^{\mu}=-1,\quad P_{\mu}S^{\mu}=0\,.
\end{equation}
Introducing a light-cone vector $n$%
\begin{equation}
n_{\mu}n^{\mu}=0\,,
\end{equation}
one can write the following decomposition \cite{BM-98}
\begin{equation*}
\frac{2(nP)}{(2\pi )^{3}M}\int d^{4}z\delta (nz)e^{ik\cdot z}\langle P,S|%
\bar{\psi}_{j}(0)\psi _{i}(z)|P,S\rangle
\end{equation*}
\begin{equation*}
=\left\{ f_{1}\frac{(P\gamma )}{M}+f_{1T}^{\perp }\,\epsilon _{\mu \nu \rho
\sigma }\gamma ^{\mu }\frac{P^{\nu }k_{T}^{\rho }S_{T}^{\sigma }}{M^{2}}%
\right.
\end{equation*}
\begin{equation*}
-\left[ \lambda \,g_{1L}-\frac{(k_TS_T)}{M}\,g_{1T}%
\right] \,\frac{(P\gamma )\gamma _{5}}{M}-h_{1T}\,\frac{i\sigma _{\mu \nu
}\gamma _{5}S_{T}^{\mu }P^{\nu }}{M}
\end{equation*}
\begin{equation}
\left. -\left[ \lambda h_{1L}^{\perp }-\frac{(k_TS_T)}{
M}\,h_{1T}^{\perp }\,\right] \,\frac{i\sigma _{\mu \nu }\gamma
_{5}k_{T}^{\mu }P^{\nu }}{M^{2}}+h_{1}^{\perp }\,\frac{\sigma _{\mu \nu
}k_{T}^{\mu }P^{\nu }}{M^{2}}\right\} _{ij}+\ldots  \label{Phi-naive}
\end{equation}%
where%
\begin{equation}
\lambda =M\frac{(nS)}{(nP)}\,.  \label{lambda-def}
\end{equation}%
The transverse components of vectors ($k_{T},S_{T}$ etc.) are defined as
projections on the plane orthogonal to $n$ and $P$. Parton distributions
appearing in the RHS of Eq. (\ref{Phi-naive}) depend on $|\mathbf{k}_{T}|^{2}$
and
\begin{equation}
x=\frac{(nk)}{(nP)}\,.
\end{equation}
The ellipsis in the RHS of Eq.~(\ref{Phi-naive}) stands for the higher twist
contributions.

Functions $f_{1T}^{\perp },h_{1}^{\perp }$ are $T$-odd and vanish in the
framework of the naive definition (\ref{Phi-naive}). However, as it was
noted in Ref. \cite{Collins-02}, the naive definition of transverse-momentum
dependent parton distributions (\ref{Phi-naive}) should be modified by
including properly chosen Wilson lines. Let us introduce the compact notation%
\begin{equation}
\Psi ^{(\pm )}(z,n)=W(\pm \infty ,z,n)\psi (z)\,,  \label{psi-plus-def}
\end{equation}%
where $W(\pm \infty ,z,n)$ is the Wilson line directed from the point $z$ to
the future (past) infinity along the light-cone vector $n$. By analogy with
Eq. (\ref{Phi-naive}) we can write the following decomposition:
\begin{equation*}
\frac{2(nP)}{(2\pi )^{3}M}\int d^{4}z\delta (nz)e^{ik\cdot z}\langle P,S|%
\bar{\Psi}_{j}^{(\pm )}(0)\Psi _{i}^{(\pm )}(z)|P,S\rangle \,,
\end{equation*}%
\begin{equation*}
=\left\{ f_{1}^{(\pm )}\frac{(P\gamma )}{M}+f_{1T}^{\perp (\pm )}\,\epsilon
_{\mu \nu \rho \sigma }\gamma ^{\mu }\frac{P^{\nu }k_{T}^{\rho
}S_{T}^{\sigma }}{M^{2}}\right.
\end{equation*}%
\begin{equation*}
-\left[ \lambda \,g_{1L}^{(\pm )}-\frac{(k_TS_T)}{M}
\,g_{1T}^{(\pm )}\right] \,\frac{(P\gamma )\gamma _{5}}{M}-h_{1T}^{(\pm )}\,%
\frac{i\sigma _{\mu \nu }\gamma _{5}S_{T}^{\mu }P^{\nu }}{M}
\end{equation*}%
\begin{equation}
\left. -\left[ \lambda h_{1L}^{\perp (\pm )}
-\frac{(k_TS_T)}{M}\,h_{1T}^{\perp (\pm )}\,\right] \,\frac{i\sigma _{\mu \nu }\gamma
_{5}k_{T}^{\mu }P^{\nu }}{M^{2}}+h_{1}^{\perp (\pm )}\,\frac{\sigma _{\mu
\nu }k_{T}^{\mu }P^{\nu }}{M^{2}}\right\} _{ij}+\ldots
\label{Phi-decomposition}
\end{equation}%
which defines functions%
\begin{equation}
F_{r}^{(\pm )}=\{f_{1}^{(\pm )},\,g_{1T}^{(\pm )},\,g_{1L}^{(\pm
)},\,f_{1T}^{\perp (\pm )},\,h_{1}^{\perp (\pm )},\,h_{1L}^{\perp (\pm
)},\,h_{1T}^{\perp (\pm )},\,h_{1T}^{(\pm )}\,\}  \label{F-a-def}
\end{equation}%
depending on $x$ and $|\mathbf{k}_{T}|^{2}$%
\begin{equation}
F_{r}^{(\pm )}=F_{r}^{(\pm )}(x,|\mathbf{k}_{T}|^{2})\,.
\end{equation}

The $T$-invariance leads to the following relations between $F_{r}^{(+)}$
and $F_{r}^{(-)}$%
\begin{equation}
F_{r}^{(+)}=\left\{ 
\begin{array}{ll}
-F_{r}^{(-)} & \mathrm{for}\,\,f_{1T}^{\perp },h_{1}^{\perp }\,, \\ 
F_{r}^{(-)} & \mathrm{otherwise.}%
\end{array}%
\right.  \label{F-T-relations}
\end{equation}%
which allows nonvanishing functions $f_{1T}^{\perp (\pm )},h_{1}^{\perp (\pm
)}$.

Multiplying Eq.~(\ref{Phi-decomposition}) by $(n\gamma )$ both on the left
and on the right we can get rid of the higher twist contributions. Taking the
nucleon at rest
\begin{equation}
P^{\mu }=(M,\mathbf{0})\,,\quad S^{\mu }=(0,\mathbf{S})\,,\quad |\mathbf{S}%
|=1\,,  \label{S-rest}
\end{equation}
\begin{equation}
\mathbf{S}=-\lambda \mathbf{n}+\mathbf{S}_{T}\,,\quad (\mathbf{S}_{T}\mathbf{%
n})=0  \label{S-lambda-S-T}
\end{equation}%
and normalizing the light-cone vector $n$ so that
\begin{equation}
n^{0}=|\mathbf{n}|=1\,,  \label{n-normalization}
\end{equation}%
we find
\begin{equation*}
\int \frac{d^{3}z}{(2\pi )^{3}}\exp \left[ ixM(\mathbf{nz})-i(\mathbf{k}_{T}%
\mathbf{z})\right]
\end{equation*}%
\begin{equation*}
\times \left. \langle P,S|\left[ \Pi (\mathbf{n})\Psi ^{(\pm )}(0,n)\right]
_{j}^{\dagger }\left[ \Pi (\mathbf{n})\Psi ^{(\pm )}(z,n)\right]
_{i}|P,S\rangle \right| _{z^{0}=n^{k}z^{k},\,\mathbf{P}=0}
\end{equation*}%
\begin{equation*}
=\frac{1}{2}\,\left\{ \left[ f_{1}^{(\pm )}+f_{1T}^{\perp (\pm )}\,\frac{1}{M%
}\left[ \mathbf{n\times k}_{T}\right] \cdot \mathbf{S}+\left( -(\mathbf{nS}%
)\,g_{1L}^{(\pm )}+\frac{(\mathbf{k}_{T}\mathbf{S})}{M}\,g_{1T}^{(\pm
)}\right) \right. \gamma _{5}\right.
\end{equation*}%
\begin{equation*}
-\left( -(\mathbf{nS})h_{1L}^{\perp (\pm )}+\frac{(\mathbf{k}_{T}\mathbf{S})%
}{M}h_{1T}^{\perp (\pm )}\right) \frac{i}{M}([\mathbf{n\times {%
\mbox{\boldmath$\gamma$}}}]\cdot \mathbf{k}_{T})
\end{equation*}%
\begin{equation}
\left. \left. -h_{1T}^{(\pm )}i([\mathbf{n\times {\mbox{\boldmath$\gamma$}}}%
]\cdot \mathbf{S)-}h_{1}^{\perp (\pm )}\,\frac{i}{M}(\mathbf{k}_{T}\mathbf{{%
\mbox{\boldmath$\gamma$}}})\right] \Pi (\mathbf{n})\right\}
_{ij}\,.  \label{st-decomp-nonrel-n}
\end{equation}%
The projector $\Pi (\mathbf{n})$ is defined as follows:%
\begin{equation}
\Pi (\mathbf{n})=\frac{1}{2}\gamma ^{0}(n\gamma )
=\frac{1-\gamma^{0}(\mathbf{n} \mbox{\boldmath$\gamma$})}{2}\,,  \label{Pi-n-def}
\end{equation}%
\begin{equation}
\left[ \Pi (\mathbf{n})\right] ^{2}=\Pi (\mathbf{n})\,.
\end{equation}%
The transverse components of vectors are orthogonal to $\mathbf{n}$:%
\begin{equation}
(\mathbf{k}_{T}\mathbf{n})=0\,.
\end{equation}

The structure of the paper is as follows.
In Section~\ref{Results-section} the main results are described.
Section~\ref{baryons-section} contains a brief summary of the general
properties of the large-$N_c$ baryons.
The  transverse-momentum dependent quark distributions
at large $N_c$ are studied in Section~\ref{distributions-section}.
Several remarks about the $T$-odd distributions are made in
Section~\ref{T-odd-section}.

\section{Results}
\label{Results-section}
\setcounter{equation}{0}

In this section the main results of the paper are described. Let us first
remind the well known large-$N_{c}$ properties of the usual ($k_{\perp}$
integrated) parton distributions. At large number of colors $N_{c}$, parton
distributions are concentrated at values of Bjorken variable $x\sim
N_{c}^{-1}$. In the naive quark model this can be qualitatively explained as
a result of the distribution of the nucleon momentum in the infinite
momentum frame between $N_{c}$ quarks. Therefore in the large-$N_{c}$ limit,
the $x$ dependence $x$ appears via the product $xN_{c}=O(N_{c}^{0})$. For
example, the twist-2 parton distributions: unpolarized $q(x)$,
longitudinally polarized $\Delta_{L}q(x)$ and transversity $\Delta_{T}q(x)$%
\begin{equation}
q_{i}=(q,\Delta_{L}q,\Delta_{T}q)
\end{equation}
have the following large-$N_{c}$ behavior%
\begin{equation}
q_{i}(x)=N_{c}^{2}\left[ f_{i}(xN_{c})+O\left( \frac{1}{N_{c}}\right) \right]
\label{q-i-f-i}
\end{equation}
where functions $f_{i}$ are $N_{c}$ independent. Concerning the isospin
properties of parton distributions, we can make the flavor decomposition
into the isoscalar ($T=0$) and isovector ($T=1$) parts:%
\begin{equation}
\left[ q_{i}(x)\right] _{t_{3},I_{3}}=q_{i}^{T=0}(x)+\mathrm{sign}%
(t_{3}I_{3})q_{i}^{T=1}(x)\,.
\end{equation}
Here $t_{3},I_{3}=\pm1/2$ are the isospin projections of the quark and
nucleon respectively. In the leading order of the $1/N_{c}$-expansion we
have the following isospin structure:
\begin{equation}
\begin{array}{ll}
q: & T=0, \\ 
\Delta_{L}q,\Delta_{T}q: & T=1\,.%
\end{array}
\label{q-i-isospin}
\end{equation}

The large-$N_{c}$ behavior (\ref{q-i-f-i}), (\ref{q-i-isospin}) was first
observed in the context of the chiral quark-soliton model \cite{DPPPW-96}
but can be also derived using the standard methods of large-$N_{c}$ counting
in QCD \cite{EGP-00}.

In the case of the transverse-momentum distributions $F_{r}^{(\pm)}$ (\ref%
{F-a-def}) we find the following large-$N_{c}$ structure:
\begin{equation}
F_{r}^{(\pm)}(x,|\mathbf{k}_{T}|^{2})=N_{c}^{m_{r}}\left[ G_{r}^{(\pm
)}(xN_{c},|\mathbf{k}_{T}|^{2})+O\left( \frac{1}{N_{c}}\right) \right] \,,
\label{F-G}
\end{equation}
where functions $G_{r}^{(\pm)}$ are $N_{c}$ independent:%
\begin{equation}
G_{r}^{(\pm)}=O(N_{c}^{0})\,.
\end{equation}
The $N_c$ orders $m_{r}$ are given in Table \ref{table-1}. Also note that in the
leading order of the large-$N_{c}$ limit every function $F_{r}$ is either
isoscalar ($T_{r}=0$) or isovector ($T_{r}=1$). This structure is also shown
in Table \ref{table-1}. We see from this table
that the distributions insensitive to the nucleon spin ($f_{1}$ and $%
h_{1}^{\perp}$) are isoscalar, whereas other distributions dependent on the
nucleon spin are isovector
\footnote{This correlation between the spin and isospin structures is a general
property of nucleon observables in the large-$N_{c}$ limit. It is a consequence
of the spin-flavor symmetry of the large-$N_c$ baryons~\cite{baryons-83}.}.

\begin{table}[ptb]
\begin{tabular}{||l|l|l||}
\hline\hline
\vphantom{\rule{0em}{1.5em}} $F_{r}^{(\pm)}$ & $m_{r}$ & $T_{r}$ \\ 
\hline\hline
\vphantom{\rule{0em}{1.5em}} $f_{1}^{(\pm)}$ & $2$ & $0$ \\ \hline
\vphantom{\rule{0em}{1.5em}} $g_{1T}^{(\pm)}$ & $3$ & $1$ \\ \hline
\vphantom{\rule{0em}{1.5em}} $g_{1L}^{(\pm)}$ & $2$ & $1$ \\ \hline
\vphantom{\rule{0em}{1.5em}} $f_{1T}^{\perp(\pm)}$ & $3$ & $1$ \\ \hline
\vphantom{\rule{0em}{1.5em}} $h_{1}^{\perp(\pm)}$ & $3$ & $0$ \\ \hline
\vphantom{\rule{0em}{1.5em}} $h_{1L}^{\perp(\pm)}$ & $3$ & $1$ \\ \hline
\vphantom{\rule{0em}{1.5em}} $h_{1T}^{\perp(\pm)}$ & $4$ & $1$ \\ \hline
\vphantom{\rule{0em}{1.5em}} $h_{1T}^{(\pm)}$ & $2$ & $1$ \\ \hline\hline
\end{tabular}%
\caption{The $N_{c}^{m_{r}}$ behavior and the dominant isospin-$T_{r}$ structure
of transverse-momentum dependent quark distributions $F_{r}^{\pm}$.}
\label{table-1}
\end{table}

Note that the nucleon mass $M=O(N_{c})$ grows with $N_{c}$. The definition
of the transverse-momentum dependent distributions (\ref{st-decomp-nonrel-n})
contains explicit factors of $M$ introduced as an auxiliary dimensional
parameter. These factors of $M$ contribute to our counting of $N_{c}$
orders. This should be kept in mind if one tries to draw phenomenological
conclusions about the suppression or enhancement of various distributions to
physical processes from the large $N_c$ counting.

\section{Large-$N_{c}$ baryons}
\label{baryons-section}

\setcounter{equation}{0}

In the large-$N_{c}$ limit the mass of the nucleon $M$ grows as $O(N_{c})$
\cite{Witten-79}:
\begin{equation}
M=O(N_{c})\,.  \label{M-Nc}
\end{equation}
In the naive quark model this can be attributed to the fact that nucleon
consists of $N_{c}$ quarks. According to the standard picture of the large-$%
N_{c}$ baryons in QCD, the lowest baryonic excitations have spin $S$ equal
to isospin $I$ (in the case of the $SU(2)$ flavor group) \cite{baryons-83} 
\begin{equation}
I=S=\frac{1}{2},\frac{3}{2},\ldots  \label{nucleon-series}
\end{equation}
These states become degenerate at large $N_{c}$
\begin{equation}
M_{B}=M\left[ 1+O(N_{c}^{-1})\right]
\end{equation}
and can be described as ``rotational excitations'' of the mean field
solution of the effective theory describing large-$N_{c}$ QCD. We use
subscript $B$%
\begin{equation}
B=(I,I_{3},S_{3}),\quad I=S
\end{equation}
for the set of discrete quantum numbers characterizing the baryon state.

It is convenient to work in the frame where this heavy large-$N_{c}$ nucleon
is nonrelativistic. But we still use the relativistic normalization of
nucleon states:

\begin{equation*}
\langle B_{1},\mathbf{P}_{1}|B_{2},\mathbf{P}_{2}\rangle=2P^{0}(2\pi
)^{3}\delta_{B_{1}B_{2}}\delta^{(3)}(\mathbf{P}_{1}-\mathbf{P}_{2})
\end{equation*}%
\begin{equation}
=2M\left[ 1+O(N_{c}^{-1})\right] \delta_{B_{1}B_{2}}(2\pi)^{3}\delta ^{(3)}(%
\mathbf{P}_{1}-\mathbf{P}_{2})\,.  \label{B-normalization}
\end{equation}

The exact form of the effective action describing large-$N_{c}$ QCD is not
known. But certain conclusions can be derived from the symmetry of the mean
field: the mean field is assumed to have the spin-flavor invariance: it is
invariant under simultaneous space and isospin rotations. In the leading
order of the $1/N_{c}$-expansion, for a certain class of gauge invariant
operators $\hat{A}$ (including bilocal quark operators accompanied by Wilson
lines) their matrix elements over baryon states (\ref{nucleon-series}) can
be represented in the form%
\begin{equation*}
\langle B_{1},\mathbf{P}_{1}|\hat{A}|B_{2},\mathbf{P}_{2}\rangle=2M\int d^{3}%
\mathbf{Xe}^{i(\mathbf{P}_{2}-\mathbf{P}_{1})\cdot\mathbf{X}}
\end{equation*}%
\begin{equation}
\times\int dR\phi_{B_{1}}^{\ast}(R)\phi_{B_{2}}(R)A(\mathbf{X},R)\left[
1+O\left( \frac{1}{N_{c}}\right) \right] \,.  \label{A-ME}
\end{equation}
Here $A(\mathbf{X},R)$ is the mean field function associated with the
operator $\hat{A}$. The 3-vector $\mathbf{X}$ and the matrix $R\in SU(2)$
have the meaning of the collective coordinates corresponding to the
3-translations and isotopic rotations respectively. The function $e^{i%
\mathbf{PX}}$ can be interpreted as the translational wave function of the
nucleon with momentum $P$, whereas $\phi_{B}(R)$ is its ``rotational'' wave
function. For the eigenstates of spin and isospin, these rotational wave
functions can be expressed in terms of Wigner functions $D_{mm'}^{j}(R)$ (see Appendix \ref{Wigner-functions-appendix}):%
\begin{equation}
\phi_{I=S,I_{3}S_{3}}(R)=(-1)^{I+I_{3}}\sqrt{2I+1}D_{-I_{3},S_{3}}^{I=S}(R)%
\,.  \label{phi-D-def}
\end{equation}
Here the normalization coefficient is fixed by the condition
\begin{equation}
\int dR\phi_{B_{1}}^{\ast}(R)\phi_{B_{2}}(R)=\delta_{B_{1}B_{2}}
\label{phi-orthonormal}
\end{equation}
and the $SU(2)$ Haar measure $dR$ is normalized as follows:
\begin{equation}
\int dR=1\,.
\end{equation}
Usually equations of the type (\ref{A-ME}) appear in various soliton models
of nucleon. In these models one can find function $A(\mathbf{X},R)$ using
the the action of the model. In the case of large-$N_{c}$ QCD the action of
the corresponding effective theory is not known. Therefore for most of
operators $\hat{A}$ the explicit form of the function $A(\mathbf{X},R)$ is
not known. However, it is still possible to derive interesting consequences
from the representation (\ref{A-ME}).

As a simple illustration of Eq. (\ref{A-ME}) one can consider the
case of the unit operator:
\begin{equation}
\hat{A}=1\,,\quad A(\mathbf{X},R)=1\,.
\end{equation}
Then using relation (\ref{phi-orthonormal}) and%
\begin{equation}
\int d^{3}\mathbf{Xe}^{i(\mathbf{P}_{2}-\mathbf{P}_{1})\cdot\mathbf{X}%
}=(2\pi)^{3}\delta^{(3)}(\mathbf{P}_{1}-\mathbf{P}_{2})
\end{equation}
we can ``derive'' the normalization condition (\ref{B-normalization}) from
Eq. (\ref{A-ME}).

\section{Parton distributions at large $N_{c}$}
\label{distributions-section}

\setcounter{equation}{0}

Let us apply the general equation (\ref{A-ME}) to the following operator $%
\hat{A}$%
\begin{equation}
\hat{A}_{s_{2}t_{2}s_{1}t_{1}}^{(\pm)}(z_{2},z_{1},n)=\left[
\Psi_{s_{1}t_{1}}^{(\pm)}(z_{1},n)\right] ^{\dagger}\Psi_{s_{2}t_{2}}^{(%
\pm)}(z_{2},n)\,.  \label{A-psi-def}
\end{equation}
Here $\Psi_{s_{2}t_{2}}^{(\pm)}$ are quark fields (\ref{psi-plus-def})
modified by the Wilson lines directed along the light-cone vector $n$. The
indices $t_{k}$ stand for the isospin, and the $s_{k}$ are Dirac 4-spinor
indices.

For this operator $\hat{A}^{(\pm)}(z_{1},z_{2},n)$ relation (\ref{A-ME})
takes the form%
\begin{equation*}
\langle B_{1},\mathbf{P}_{1}|\left[ \Psi_{s_{1}t_{1}}^{(\pm)}(z_{1},n)\right]
^{\dagger}\Psi_{s_{2}t_{2}}^{(\pm)}(z_{2},n)|B_{2},\mathbf{P}_{2}\rangle=2M%
\left[ 1+O(N_{c}^{-1})\right]
\end{equation*}%
\begin{equation}
\times\int d^{3}\mathbf{Xe}^{i(\mathbf{P}_{2}-\mathbf{P}_{1})\cdot\mathbf{X}%
}\int dR\phi_{B_{1}}^{\ast}(R)\phi_{B_{2}}(R)A_{s_{2}t_{2}s_{1}t_{1}}^{(\pm
)}(z_{2},z_{1},n|\mathbf{X},R)\,.  \label{A-ME-2}
\end{equation}
Starting from this expression we can compute the matrix element appearing
the LHS of Eq.~(\ref{st-decomp-nonrel-n}):%
\begin{equation*}
\int\frac{d^{3}z}{(2\pi)^{3}}\exp\left[ ixM(\mathbf{nz})-i(\mathbf{k}_{T}%
\mathbf{z})\right]
\end{equation*}%
\begin{equation*}
\times\langle B_{1},\mathbf{P}=0|\left[ \Pi(\mathbf{n})\Psi^{(\pm )}(0,n)%
\right] _{s_{1}t_{1}}^{\dagger}\left[ \Pi(\mathbf{n})\Psi^{(\pm )}((\mathbf{%
nz}),\mathbf{z};n)\right] _{s_{2}t_{2}}|B_{2},\mathbf{P}=0\rangle
\end{equation*}%
\begin{equation}
=2MN_{c}\left[ 1+O(N_{c}^{-1})\right] \int
dR\phi_{B_{1}}^{\ast}(R)\phi_{B_{2}}(R)\left[ R_{t_{2}t_{2}^{\prime}}%
\rho_{s_{2}t_{2}^{\prime },s_{1}t_{1}^{\prime}}^{(\pm)}(x,\mathbf{k}_{T},%
\mathbf{n})(R^{-1})_{t_{1}^{\prime}t_{1}}\right] \,.  \label{psi-psi-MF}
\end{equation}
This equation is derived in Appendix \ref{A-rho-appendix} where one can find
the explicit expression for the function $\rho^{(\pm)}$ in terms of $%
A^{(\pm)}$. Actually neither $A^{(\pm)}$ nor $\rho^{(\pm)}$ can be computed
from the first principles, since large-$N_{c}$ QCD is not solved. Therefore
we can proceed relying only on the symmetries of $\rho^{(\pm)}$.

Using the $P$-invariance properties of the matrix element appearing in the
LHS of Eq.~(\ref{psi-psi-MF}) we find
\begin{equation}
\gamma^{0}\rho^{(\pm)}(x,\mathbf{k}_{T},\mathbf{n})\gamma^{0}=\rho^{(\pm
)}(x,-\mathbf{k}_{T},-\mathbf{n})\,.  \label{rho-P-symmetry}
\end{equation}
Similarly the $T$-invariance leads to the following relation%
\begin{equation}
\left[ \rho^{(+)}(x,\mathbf{k}_{T},\mathbf{n})\right] ^{\mathrm{tr}}=\tau^{2}%
\mathcal{C}\gamma_{5}\rho^{(-)}(x,-\mathbf{k}_{T},-\mathbf{n})\gamma_{5}%
\mathcal{C}^{-1}\tau^{2}\,.  \label{rho-T-invariance}
\end{equation}
Here the superscript \emph{tr} stands for the matrix transposition (both in
spin and flavor indices) and $\mathcal{C}$ is the matrix with the property%
\begin{equation}
\mathcal{C}\gamma_{\mu}\mathcal{C}^{-1}=-\gamma_{\mu}^{\mathrm{tr}}\,.
\end{equation}
We use notation $\tau^{a}$ for the isospin Pauli matrices and take into
account that%
\begin{equation}
\tau^{2}\tau^{a}\tau^{2}=-(\tau^{a})^{\mathrm{tr}}\,.
\end{equation}

Due to the presence of the projector $\Pi(\mathbf{n})$ (\ref{Pi-n-def}) in
the LHS of Eq.~(\ref{psi-psi-MF}) we have%
\begin{equation}
\Pi(\mathbf{n})\rho^{(\pm)}(x,\mathbf{k}_{T},\mathbf{n})\Pi(\mathbf{n}%
)=\rho^{(\pm)}(x,\mathbf{k}_{T},\mathbf{n})\,.  \label{Pi-rho-constraint}
\end{equation}
In Appendix \ref{rho-structures-appendix} we find the general solution of
constraints (\ref{rho-P-symmetry}) and (\ref{Pi-rho-constraint}):

\begin{equation*}
\rho ^{(\pm )}(x,\mathbf{k}_{T},\mathbf{n})=\left\{ V_{1}^{(\pm
)}+V_{2}^{(\pm )}\gamma _{5}(\mathbf{k}_{T}\mathbf{{\mbox{\boldmath$\tau$}}}%
)+V_{3}^{(\pm )}\gamma _{5}(\mathbf{n{\mbox{\boldmath$\tau$}}})+V_{4}^{(\pm
)}\left( \mathbf{n}\cdot \lbrack \mathbf{k}_{T}\times \mathbf{{%
\mbox{\boldmath$\tau$}}}]\right) \right. 
\end{equation*}%
\begin{equation*}
+V_{5}^{(\pm )}i(\mathbf{k}_{T}\mathbf{{\mbox{\boldmath$\gamma$}}}%
)+V_{6}^{(\pm )}i\left( \mathbf{n}\cdot \lbrack \mathbf{{\mbox{\boldmath$%
\gamma$}}}\times \mathbf{k}_{T}]\right) (\mathbf{n{\mbox{\boldmath$\tau$}}})
\end{equation*}%
\begin{equation}
\left. +V_{7}^{(\pm )}i\left( \mathbf{n}\cdot \lbrack \mathbf{{%
\mbox{\boldmath$\gamma$}}}\times \mathbf{k}_{T}]\right) (\mathbf{k}_{T}%
\mathbf{{\mbox{\boldmath$\tau$}}})+V_{8}^{(\pm )}i\left( \mathbf{n}\cdot
\lbrack \mathbf{{\mbox{\boldmath$\gamma$}}}\times \mathbf{{%
\mbox{\boldmath$\tau$}}}]\right) \right\} \Pi (\mathbf{n})\,.  \label{rho-A}
\end{equation}%
Here%
\begin{equation}
V_{r}^{(\pm )}=V_{r}^{(\pm )}(x,|\mathbf{k}_{T}|^{2})\,.  \label{V-r-scalar}
\end{equation}%
According to Eqs.~(\ref{rho-A}) and (\ref{rho-Nc}) we have%
\begin{equation}
V_{r}=O(N_{c}^{0})\,.  \label{V-N-c}
\end{equation}%
The $T$-invariance (\ref{rho-T-invariance}) leads to the property%
\begin{equation}
V_{r}^{(+)}=\left\{ 
\begin{array}{ll}
-V_{r}^{(-)} & \mathrm{for}\,\,r=4,5, \\ 
V_{r}^{(-)} & \mathrm{otherwise\,.}%
\end{array}%
\right.   \label{V-T-parity}
\end{equation}%
In Appendix \ref{R-integrals-appendix}, using decomposition (\ref{rho-A}),
we compute the integral over $R$ appearing in the RHS of Eq.~(\ref%
{psi-psi-MF}):%
\begin{equation*}
\int dR\phi _{I_{3},\mathbf{S}}^{\ast }(R)\phi _{I_{3}\mathbf{,S}}\left[
R_{t_{2}t_{2}^{\prime }}\rho _{s_{2}t_{2}^{\prime },s_{1}t_{1}^{\prime
}}^{(\pm )}(x,\mathbf{k}_{T},\mathbf{n})(R^{-1})_{t_{1}^{\prime }t_{1}}%
\right] 
\end{equation*}%
\begin{equation*}
=\left( \delta _{t_{2}t_{1}}\left[ V_{1}^{(\pm )}+V_{5}^{(\pm )}i(\mathbf{k}%
_{T}\mathbf{{\mbox{\boldmath$\gamma$}}})\right] _{s_{2}s_{1}}\right. 
\end{equation*}%
\begin{equation*}
-\frac{2}{3}I_{3}(\tau ^{3})_{t_{2}t_{1}}\left\{ V_{2}^{(\pm )}\gamma _{5}(%
\mathbf{S}\mathbf{k}_{T})+V_{3}^{(\pm )}(\mathbf{nS}) \gamma _{5}\right. 
\end{equation*}%
\begin{equation*}
+V_{4}^{(\pm )}\left[ \mathbf{n\times k}_{T}\right] \cdot \mathbf{S}%
+V_{6}^{(\pm )}(\mathbf{nS}) i\left( \mathbf{n}\cdot \lbrack \mathbf{{%
\mbox{\boldmath$\gamma$}}}\times \mathbf{k}_{T}]\right) 
\end{equation*}%
\begin{equation}
\left. \left. +V_{7}^{(\pm )}i\left( \mathbf{n}\cdot \lbrack \mathbf{{%
\mbox{\boldmath$\gamma$}}}\times \mathbf{k}_{T}]\right) (\mathbf{S}%
\mathbf{k}_{T})+V_{8}^{(\pm )}i\left( \mathbf{n\times {\mbox{\boldmath$%
\gamma$}}}\right) \cdot \mathbf{S}\right\} \Pi (\mathbf{n})\right)
_{s_{2}s_{1}}\,.  \label{int-R-res}
\end{equation}%
Inserting this expression into Eq.~(\ref{psi-psi-MF}) and combining the
result with Eq.~(\ref{st-decomp-nonrel-n}) we find%
\begin{equation*}
\delta _{t_{2}t_{1}}\left[ V_{1}^{(\pm )}+V_{5}^{(\pm )}i(\mathbf{k}_{T}%
\mathbf{{\mbox{\boldmath$\gamma$}}})\right] _{s_{2}s_{1}}
\end{equation*}%
\begin{equation*}
-\frac{2}{3}I_{3}(\tau ^{3})_{t_{2}t_{1}}\left\{ V_{2}^{(\pm )}\gamma _{5}(%
\mathbf{S}\mathbf{k}_{T})+V_{3}^{(\pm )}(\mathbf{nS}) \gamma _{5}+V_{4}^{(\pm
)}\left[ \mathbf{n\times k}_{T}\right] \cdot \mathbf{S}\right. 
\end{equation*}%
\begin{equation*}
\left. +V_{6}^{(\pm )}(\mathbf{nS}) i\left( \mathbf{n}\cdot \lbrack \mathbf{{%
\mbox{\boldmath$\gamma$}}}\times \mathbf{k}_{T}]\right) +V_{7}^{(\pm
)}i\left( \mathbf{n}\cdot \lbrack \mathbf{{\mbox{\boldmath$\gamma$}}}\times 
\mathbf{k}_{T}]\right) (\mathbf{S}\mathbf{k}_{T})+V_{8}^{(\pm )}i\left( 
\mathbf{n\times {\mbox{\boldmath$\gamma$}}}\right) \cdot \mathbf{S}%
\right\} _{s_{2}s_{1}}
\end{equation*}%
\begin{equation*}
=\frac{1}{4MN_{c}}\,\left\{ f_{1}^{(\pm )}+f_{1T}^{\perp (\pm )}\,\frac{1}{M}%
\left[ \mathbf{n\times k}_{T}\right] \cdot \mathbf{S}\right. +\left[ -(%
\mathbf{nS})\,g_{1L}^{(\pm )}+\frac{(\mathbf{k}_{T} \mathbf{S})}{M}%
\,g_{1T}^{(\pm )}\right] \gamma _{5}
\end{equation*}%
\begin{equation*}
-\left[ -(\mathbf{nS})h_{1L}^{\perp (\pm )}+\frac{(\mathbf{k}_{T} 
\mathbf{S})}{M}h_{1T}^{\perp (\pm )}\right] \frac{i}{M}([\mathbf{n\times {%
\mbox{\boldmath$\gamma$}}}]\cdot \mathbf{k}_{T})
\end{equation*}%
\begin{equation}
\left. -h_{1T}^{(\pm )}i([\mathbf{n\times {\mbox{\boldmath$\gamma$}}}]\cdot 
\mathbf{S)-}h_{1}^{\perp (\pm )}\,\frac{i}{M}(\mathbf{k}_{T}\mathbf{{%
\mbox{\boldmath$\gamma$}}})\right\} _{s_{2}t_{2},s_{1}t_{1}}\,.
\label{V-F-res}
\end{equation}%
Using Eq. (\ref{S-lambda-S-T}), we see that the LHS of Eq. (\ref{V-F-res})
contains the same spin-isospin structures as the RHS so that

\begin{equation}
\left( f_{1}^{(\pm )}\right) _{t_{1}t_{2}}=4MN_{c}\delta
_{t_{2}t_{1}}V_{1}^{(\pm )}=O(N_{c}^{2})\,,  \label{V-F-1}
\end{equation}%
\begin{equation}
\left( g_{1T}^{(\pm )}\right) _{t_{1}t_{2}}=-\frac{8}{3}I_{3}(\tau
^{3})_{t_{2}t_{1}}M^{2}N_{c}V_{2}^{(\pm
)}=O(N_{c}^{3})\,,
\end{equation}%
\begin{equation}
\left( g_{1L}^{(\pm )}\right) _{t_{1}t_{2}}=\frac{8}{3}I_{3}(\tau
^{3})_{t_{2}t_{1}}MN_{c}V_{3}^{(\pm )}=O(N_{c}^{2})\,,
\end{equation}%
\begin{equation}
\left( f_{1T}^{\perp (\pm )}\right) _{t_{1}t_{2}}=-\frac{8}{3}I_{3}(\tau
^{3})_{t_{2}t_{1}}M^{2}N_{c}V_{4}^{(\pm
)}=O(N_{c}^{3})\,,
\end{equation}%
\begin{equation}
\left( h_{1}^{\perp }\right) _{t_{1}t_{2}}=-4M^{2}N_{c}\delta
_{t_{2}t_{1}}V_{5}^{(\pm )}=O(N_{c}^{3})\,,
\end{equation}%
\begin{equation}
\left( h_{1L}^{\perp (\pm )}\right) _{t_{1}t_{2}}=-\frac{8}{3}I_{3}(\tau
^{3})_{t_{2}t_{1}}M^{2}N_{c}V_{6}^{(\pm
)}=O(N_{c}^{3})\,,
\end{equation}%
\begin{equation}
\left( h_{1T}^{\perp (\pm )}\right) _{t_{1}t_{2}}=\frac{8}{3}I_{3}(\tau
^{3})_{t_{2}t_{1}}M^{3}N_{c}V_{7}^{(\pm
)}=O(N_{c}^{4})\,,
\end{equation}%
\begin{equation}
\left( h_{1T}^{(\pm )}\right) _{t_{1}t_{2}}=\frac{8}{3}I_{3}(\tau
^{3})_{t_{2}t_{1}}MN_{c}V_{8}^{(\pm )}=O(N_{c}^{2})\,.
\label{V-F-8}
\end{equation}
The orders of $N_{c}$ are determined taking into account Eqs. (\ref{M-Nc})
and (\ref{V-N-c}). We remind that indices $t_{1},t_{2}$ are isospin
projections of the quark fields in the definition of the parton
distributions (\ref{Phi-decomposition}). We see that distributions $f_{1}$, $%
h_{1}^{\perp }$ are isoscalar in the leading order of the $1/N_{c}$%
-expansion whereas the leading parts of all other distributions are
isovector.

Note that relations (\ref{V-F-1}) -- (\ref{V-F-8}) are compatible
with the $T$-parity properties of parton distributions (\ref{F-T-relations})
and functions $V_r$ (\ref{V-T-parity}).

Our results (\ref{V-F-1}) -- (\ref{V-F-8}) are summarized in Table \ref%
{table-1} of Section \ref{Results-section}.

\section{$T$-odd distributions in large-$N_{c}$ QCD}
\label{T-odd-section}
\setcounter{equation}{0}

The status of the $T$-odd distributions $f_{1T}^{\perp}\,$and $h_{1}^{\perp}$
was clarified in Ref. \cite{Collins-02} where it was noted that the
orientation of the Wilson lines accompanying the quark fields is sensitive
to the considered process. Therefore one deals with two sets of
distributions $F_{r}^{(+)}$ and $F_{r}^{(-)}$ which are related by the $T$%
-parity relations (\ref{F-T-relations}). In particular, for the $T$-odd
functions one has%
\begin{equation}
f_{1T}^{\perp(+)}=-f_{1T}^{\perp(-)}\,,\quad
h_{1}^{\perp(+)}=-h_{1}^{\perp(-)}\,.
\end{equation}
Due the dependence of parton distributions on the direction of Wilson lines
functions $f_{1T}^{\perp(\pm)}$, $h_{1}^{\perp(\pm)}$ do not vanish.

In our construction of the $1/N_{c}$-expansion the sensitivity of parton
distributions to the direction of Wilson lines is properly take into
account. Therefore the nonvanishing $T$-odd functions $f_{1T}^{\perp(\pm)}$, 
$h_{1}^{\perp(\pm)}$ naturally appear via the $T$-odd structures $%
V_{4},V_{5} $ (\ref{rho-A}) in our analysis of the large-$N_{c}$ limit.

As it is well known, various chiral models properly reproduce the large-$%
N_{c}$ counting of QCD. For example, the large-$N_{c}$ structure (\ref%
{q-i-f-i}), (\ref{q-i-isospin}) of $k_{\perp}$ integrated distributions
historically was first established in the context of the chiral quark
soliton model. The large-$N_{c}$ properties of the $T$-even distributions
can be also reproduced in chiral models. However, with the $T$-odd
distributions the situation is different. In Ref. \cite{ABDM-02} an attempt
was made to generate $T$-odd distributions in the chiral sigma model model.
But the $T$-odd distributions do not vanish in QCD only due to the special
choice of Wilson lines. The absence of Wilson lines (or other effective
fields imitating Wilson lines) in the chiral models does not allow to obtain 
$T$-odd distribution functions in these models \cite{PP-02}. This failure of
the chiral models has nothing to do with our analysis of large-$N_{c}$ limit
of QCD where the $T$-odd functions are clearly seen in Eq.~(\ref{V-F-res}).

\section{Conclusions}
\setcounter{equation}{0}

In this paper we have performed the large-$N_{c}$ analysis of the
transverse-momentum dependent quark distributions. The leading powers of $%
N_{c}$ are determined for all twist-2 distributions. The isospin structure
of the leading contributions is also found. We have explicitly checked that
the large-$N_{c}$ limit allows the existence of nonzero $T$-odd
distributions. The results are based only on the spin-flavor symmetry of the
large-$N_{c}$ baryons in QCD without any model assumptions.

\textbf{Acknowledgments.} The author appreciates discussions with Ya.I.
Azimov, D. Boer, A.V. Efremov, K. Goeke, A. Metz, P.J. Mulders, M.V.
Polyakov and P. Schweitzer. This work was supported by DFG and BMBF.

\renewcommand{\theequation}{\Alph{section}.\arabic{equation}} \appendix     

\section*{Appendices}

\section{Wigner functions}

\label{Wigner-functions-appendix}

Wigner functions $D_{mm^{\prime }}^{j}(R)$ correspond to the spin-$j$
representation of $SU(2)$ matrices $R$%
\begin{equation}
\sum\limits_{m^{\prime }}D_{mm^{\prime }}^{j}(R_{1})D_{m^{\prime }m^{\prime
\prime }}^{j}(R_{2})=D_{mm^{\prime \prime }}^{j}(R_{1}R_{2})\,.
\end{equation}%
Assuming the unit normalization of the $SU(2)$ Haar measure%
\begin{equation}
\int dR=1\,,
\end{equation}%
we have%
\begin{equation}
\int dR\,D_{m_{1}m_{1}^{\prime }}^{j_{1}\ast }(R)D_{m_{2}m_{2}^{\prime
}}^{j_{2}}(R)=\frac{1}{2j_{1}+1}\delta _{j_{1}j_{2}}\delta
_{m_{1}m_{2}}\delta _{m_{1}m_{2}}.
\end{equation}%
The integrals containing products of three $D$ functions can be expressed in
terms of $3j$-symbols:%
\begin{equation*}
\int dR\,D_{m_{1}m_{1}^{\prime }}^{j_{1}}(R)D_{m_{2}m_{2}^{\prime
}}^{j_{2}}(R)D_{m_{3}m_{3}^{\prime }}^{j_{3}}(R)
\end{equation*}%
\begin{equation}
=\left( 
\begin{array}{ccc}
j_{1} & j_{2} & j_{3} \\ 
m_{1} & m_{2} & m_{3}%
\end{array}%
\right) \left( 
\begin{array}{ccc}
j_{1} & j_{2} & j_{3} \\ 
m_{1}^{\prime } & m_{2}^{\prime } & m_{3}^{\prime }%
\end{array}%
\right) \,.
\end{equation}

\section{Large-$N_{c}$ limit and spin-flavor symmetry}

\label{A-rho-appendix} \setcounter{equation}{0}

In this appendix we derive relation (\ref{psi-psi-MF}). As it was mentioned
above, we cannot compute functions $A^{(\pm )}$ appearing in the RHS of Eq. (%
\ref{A-ME-2}) from the first principles in large-$N_{c}$ QCD. However, we
can study the symmetry properties of $A^{(\pm )}$ following from the
symmetries of the mean field describing the baryons in the (unknown)
effective theory equivalent to large-$N_{c}$ QCD.

1) The mean field is ``static'' (i.e. translationally invariant in time):%
\begin{equation}
A_{s_{2}t_{2}s_{1}t_{1}}^{(\pm )}(z_{2}^{0}+T,\mathbf{z}_{2};z_{1}^{0}+T,%
\mathbf{z}_{1};n|\mathbf{X},R)=A_{s_{2}t_{2}s_{1}t_{1}}^{(\pm )}(z_{2}^{0},%
\mathbf{z}_{2};z_{1}^{0},\mathbf{z}_{1};n|\mathbf{X},R)\,.  \label{A-static}
\end{equation}

2) The mean field has the spin-flavor symmetry \cite{baryons-83}:%
\begin{equation*}
S_{s_{2}s_{2}^{\prime }}(R)R_{t_{2}t_{2}^{\prime }}A_{s_{2}^{\prime
}t_{2}^{\prime }s_{1}^{\prime }t_{1}^{\prime }}^{(\pm )}(z_{2}^{0},O(R^{-1})%
\mathbf{z}_{2};z_{1}^{0},O(R^{-1})\mathbf{z}_{1};n^{0},O(R^{-1})\mathbf{n}|%
\mathbf{0},1)
\end{equation*}%
\begin{equation}
\times (R^{-1})_{t_{1}^{\prime }t_{1}}S_{s_{1}^{\prime
}s_{1}}(R^{-1})=A_{s_{2}t_{2}s_{1}t_{1}}^{(\pm )}(z_{2}^{0},\mathbf{z}%
_{2};z_{1}^{0},\mathbf{z}_{1};n^{0},\mathbf{n}|\mathbf{0},1)\,.
\label{A-sf-sym}
\end{equation}%
Here $R$ is an arbitrary $SU(2)$ matrix acting on the isospin indices of $%
A^{(\pm )}$. Next, $O(R)$ is the space rotation $SO(3)$ matrix associated
with the $SU(2)$ matrix $R$:%
\begin{equation}
O^{ab}(R)=\frac{1}{2}\mathrm{Tr}(\tau ^{a}R\tau ^{b}R^{-1})\,.
\label{O-ab-def}
\end{equation}%
The matrix $S(R)$ is the representation of the 3-dimensional rotations for
Dirac spinors: if 
\begin{equation}
R=\exp (i\omega ^{a}\tau ^{a})\,
\end{equation}%
then%
\begin{equation}
S(R)=\exp (i\omega ^{a}\gamma ^{5}\gamma ^{0}\gamma ^{a})\,.
\end{equation}

3) The mean field describing the nucleon is not translationally invariant
but the collective coordinate $X$ allows us to restore this invariance:

\begin{equation}
A_{s_{2}t_{2}s_{1}t_{1}}^{(\pm )}(z_{2}^{0},\mathbf{z}_{2};z_{1}^{0},\mathbf{%
z}_{1};n|\mathbf{X},R)=A_{s_{2}t_{2}s_{1}t_{1}}^{(\pm )}(z_{2}^{0},\mathbf{z}%
_{2}-\mathbf{X};z_{1}^{0},\mathbf{z}_{1}-\mathbf{X};n|\mathbf{0},R)\,.
\label{A-X-shift}
\end{equation}

4) Similarly the isotopic invariance broken by the mean field is restored by
using the collective coordinate $R$%
\begin{equation}
A_{s_{2}t_{2}s_{1}t_{1}}^{(\pm )}(z_{2}^{0},\mathbf{z}_{2};z_{1}^{0},\mathbf{%
z}_{1};n|\mathbf{X},R)=R_{t_{2}t_{2}^{\prime }}A_{s_{2}t_{2}^{\prime
}s_{1}t_{1}^{\prime }}^{(\pm )}(z_{2}^{0},\mathbf{z}_{2};z_{1}^{0},\mathbf{z}%
_{1};n|\mathbf{X},1)(R^{-1})_{t_{1}^{\prime }t_{1}}\,.  \label{A-R-rotation}
\end{equation}

Combining the properties (\ref{A-static}), (\ref{A-X-shift}), (\ref%
{A-R-rotation}), we conclude that%
\begin{equation*}
A_{s_{2}t_{2}s_{1}t_{1}}^{(\pm )}(z_{2}^{0},\mathbf{z}_{2};z_{1}^{0},\mathbf{%
z}_{1};n|\mathbf{X},R)
\end{equation*}%
\begin{equation}
=R_{t_{2}t_{2}^{\prime }}A_{s_{2}t_{2}^{\prime }s_{1}t_{1}^{\prime }}^{(\pm
)}(z_{2}^{0}-z_{1}^{0},\mathbf{z}_{2}-\mathbf{X};0,\mathbf{z}_{1}-\mathbf{X}%
;n|\mathbf{0},1)(R^{-1})_{t_{1}^{\prime }t_{1}}\,.  \label{A-X-R-A}
\end{equation}%
Let us introduce the following notation:
\begin{equation}
F_{s_{2}t_{2}^{\prime }s_{1}t_{1}^{\prime }}^{(\pm )}(T,\mathbf{z}_{2},%
\mathbf{z}_{1},n)\equiv N_{c}^{-1}A_{s_{2}t_{2}^{\prime }s_{1}t_{1}^{\prime
}}^{(\pm )}(T,\mathbf{z}_{2};0,\mathbf{z}_{1};n|\mathbf{0},1)\,.
\end{equation}%
Here we have included the factor of $N_{c}^{-1}$ into the definition of $%
F^{(\pm )}$ in order to have%
\begin{equation}
F_{s_{2}t_{2}^{\prime }s_{1}t_{1}^{\prime }}^{(\pm )}=O(N_{c}^{0})\,.
\label{F-Nc}
\end{equation}%
Indeed, the definition of the operator $\hat{A}_{s_{2}t_{2}s_{1}t_{1}}^{(\pm
)}$ (\ref{A-psi-def}) implicitly contains the contraction of the color
indices of the quark fields $\Psi ^{(\pm )}$ which leads to the $O(N_{c})$
behavior of $A_{s_{2}t_{2}s_{1}t_{1}}^{(\pm )}$.

We find from Eq.~(\ref{A-X-R-A})%
\begin{equation*}
A_{s_{2}t_{2}s_{1}t_{1}}^{(\pm )}(z_{2}^{0},\mathbf{z}_{2};z_{1}^{0},\mathbf{%
z}_{1};n|\mathbf{X},R)
\end{equation*}%
\begin{equation}
=N_{c}R_{t_{2}t_{2}^{\prime }}F_{s_{2}t_{2}^{\prime }s_{1}t_{1}^{\prime
}}^{(\pm )}(z_{2}^{0}-z_{1}^{0},\mathbf{z}_{2}-\mathbf{X},\mathbf{z}_{1}-%
\mathbf{X},n)(R^{-1})_{t_{1}^{\prime }t_{1}}\,.  \label{A-F}
\end{equation}%
and the spin-flavor symmetry relation (\ref{A-sf-sym}) takes the form%
\begin{equation*}
S_{s_{2}s_{2}^{\prime }}(R)R_{t_{2}t_{2}^{\prime }}F_{s_{2}^{\prime
}t_{2}^{\prime },s_{1}^{\prime }t_{1}^{\prime }}^{(\pm
)}(z_{2}^{0}-z_{1}^{0},O(R^{-1})\mathbf{z}_{2},O(R^{-1})\mathbf{z}%
_{1},n^{0},O(R^{-1})\mathbf{n})
\end{equation*}%
\begin{equation}
\times (R^{-1})_{t_{1}^{\prime }t_{1}}S_{s_{1}^{\prime
}s_{1}}(R^{-1})=F_{s_{2}t_{2},s_{1}t_{1}}^{(\pm )}(z_{2}^{0}-z_{1}^{0},%
\mathbf{z}_{2},\mathbf{z}_{1},n^{0},\mathbf{n})\,.  \label{F-SF-symmetry}
\end{equation}

Inserting Eqs.~(\ref{A-psi-def}), (\ref{A-F}) into Eq.~(\ref{A-ME-2}) we find

\begin{equation*}
\langle B_{1},\mathbf{P}_{1}|\left[ \Psi _{s_{1}t_{1}}^{(\pm )}(z_{1}^{0},%
\mathbf{z}_{1},n)\right] ^{\dagger }\Psi _{s_{2}t_{2}}^{(\pm )}(z_{2}^{0},%
\mathbf{z}_{2},n)|B_{2},\mathbf{P}_{2}\rangle
\end{equation*}%
\begin{equation*}
=2MN_{c}\int d^{3}\mathbf{Xe}^{i(\mathbf{P}_{2}-\mathbf{P}_{1})\cdot \mathbf{%
X}}\int dR\phi _{B_{1}}^{\ast }(R)\phi _{B_{2}}(R)
\end{equation*}%
\begin{equation}
\times \left[ R_{t_{2}t_{2}^{\prime }}F_{s_{2}t_{2}^{\prime
},s_{1}t_{1}^{\prime }}^{(\pm )}(z_{2}^{0}-z_{1}^{0},\mathbf{z}_{1}-\mathbf{X%
},\mathbf{z}_{2}-\mathbf{X},n^{0},\mathbf{n})(R^{-1})_{t_{1}^{\prime }t_{1}}%
\right] \,.
\end{equation}%
Setting here%
\begin{equation}
z_{1}=0,\quad z_{2}=z\,,\quad \mathbf{P}_{1}=\mathbf{P}_{2}=\mathbf{0},
\end{equation}%
we arrive at%
\begin{equation*}
\langle B_{1},\mathbf{P=0}|\left[ \Psi _{s_{1}t_{1}}^{(\pm )}(0,n)\right]
^{\dagger }\Psi _{s_{2}t_{2}}^{(\pm )}(z^{0},\mathbf{z};n)|B_{2},\mathbf{P=0}%
\rangle =2MN_{c}\int d^{3}\mathbf{X}
\end{equation*}%
\begin{equation}
\times \int dR\phi _{B_{1}}^{\ast }(R)\phi _{B_{2}}(R)\left[
R_{t_{2}t_{2}^{\prime }}F_{s_{2}t_{2}^{\prime },s_{1}t_{1}^{\prime }}^{(\pm
)}(z^{0},-\mathbf{X},\mathbf{z}-\mathbf{X},n^{0},\mathbf{n}%
)(R^{-1})_{t_{1}^{\prime }t_{1}}\right] \,.
\end{equation}%
Taking $z^{0}=(\mathbf{nz})$, we obtain%
\begin{equation*}
\int \frac{d^{3}z}{(2\pi )^{3}}\exp \left[ ixM(\mathbf{nz})-i(\mathbf{k}_{T}%
\mathbf{z})\right]
\end{equation*}%
\begin{equation*}
\times \langle B_{1},\mathbf{P=0}|\left[ \Psi _{s_{1}t_{1}}^{(\pm )}(0,n)%
\right] ^{\dagger }\Psi _{s_{2}t_{2}}^{(\pm )}((\mathbf{nz}),\mathbf{z}%
,n)|B_{2},\mathbf{P=0}\rangle
\end{equation*}%
\begin{equation*}
=2MN_{c}\int \frac{d^{3}z}{(2\pi )^{3}}\exp \left[ ixM(\mathbf{nz})-i(%
\mathbf{k}_{T}\mathbf{z})\right] \int d^{3}\mathbf{X}
\end{equation*}%
\begin{equation}
\times \int dR\phi _{B_{1}}^{\ast }(R)\phi _{B_{2}}(R)\left[
R_{t_{2}t_{2}^{\prime }}F_{s_{2}t_{2}^{\prime },s_{1}t_{1}^{\prime }}^{(\pm
)}((\mathbf{nz}),-\mathbf{X},\mathbf{z}-\mathbf{X},n^{0},\mathbf{n}%
)(R^{-1})_{t_{1}^{\prime }t_{1}}\right] \,.  \label{ME-0}
\end{equation}%
Defining%
\begin{equation*}
\tilde{\rho}_{s_{2}t_{2},s_{1}t_{1}}^{(\pm )}(x,\mathbf{k}_{T},\mathbf{n}%
)=\int \frac{d^{3}z}{(2\pi )^{3}}\exp \left[ ixM(\mathbf{nz})-i(\mathbf{k}%
_{T}\mathbf{z})\right]
\end{equation*}%
\begin{equation}
\times \int d^{3}\mathbf{X}F_{s_{2}t_{2},s_{1}t_{1}}^{(\pm )}((\mathbf{nz}),-%
\mathbf{X},\mathbf{z}-\mathbf{X},n^{0},\mathbf{n})\,.
\end{equation}%
we find from Eq.~(\ref{ME-0})%
\begin{equation*}
\int \frac{d^{3}z}{(2\pi )^{3}}\exp \left[ ixM(\mathbf{nz})-i(\mathbf{k}_{T}%
\mathbf{z})\right]
\end{equation*}%
\begin{equation*}
\times \langle B_{1},\mathbf{P}=0|\left[ \Psi _{s_{1}t_{1}}^{(\pm )}(0,n)%
\right] ^{(\pm )}\Psi _{s_{2}t_{2}}^{(\pm )}((\mathbf{nz}),\mathbf{z}%
;n)|B_{2},\mathbf{P}=0\rangle
\end{equation*}%
\begin{equation}
=2MN_{c}\int dR\phi _{B_{1}}^{\ast }(R)\phi _{B_{2}}(R)\left[
R_{t_{2}t_{2}^{\prime }}\tilde{\rho}_{s_{2}t_{2}^{\prime
},s_{1}t_{1}^{\prime }}^{(\pm )}(x,\mathbf{k}_{T},\mathbf{n}%
)(R^{-1})_{t_{1}^{\prime }t_{1}}\right] \,.  \label{ME-1}
\end{equation}

The $SU(2)$ spin-flavor symmetry (\ref{F-SF-symmetry}) of $%
F_{s_{2}t_{2},s_{1}t_{1}}$ results in the spin-flavor symmetry of $\tilde{%
\rho}_{s_{2}t_{2},s_{1}t_{1}}$:%
\begin{equation*}
S_{s_{2}s_{2}^{\prime }}(R)R_{t_{2}t_{2}^{\prime }}\tilde{\rho}%
_{s_{2}^{\prime }t_{2}^{\prime },s_{1}^{\prime }t_{1}^{\prime }}^{(\pm
)}(x,O(R^{-1})\mathbf{k}_{T},O(R^{-1})\mathbf{n})(R^{-1})_{t_{1}^{\prime
}t_{1}}S_{s_{1}^{\prime }s_{1}}(R^{-1})
\end{equation*}%
\begin{equation}
=\tilde{\rho}_{s_{2}t_{2},s_{1}t_{1}}^{(\pm )}(x,\mathbf{k}_{T},\mathbf{n}%
)\,.  \label{rho-tilde-SFS}
\end{equation}%
Next we introduce notation%
\begin{equation}
\rho ^{(\pm )}(x,\mathbf{k}_{T},\mathbf{n})=\Pi (\mathbf{n})\tilde{\rho}%
^{(\pm )}(x,\mathbf{k}_{T},\mathbf{n})\Pi (\mathbf{n})  \label{rho-def-Pi}
\end{equation}%
with $\Pi (\mathbf{n})$ given by (\ref{Pi-n-def}). According to Eq. (\ref%
{F-Nc}) we have%
\begin{equation}
\rho _{s_{2}t_{2},s_{1}t_{1}}^{(\pm )}=O(N_{c}^{0})\,.  \label{rho-Nc}
\end{equation}%
Eq.~(\ref{rho-tilde-SFS}) yields

\begin{equation*}
S_{s_{2}s_{2}^{\prime }}(R)R_{t_{2}t_{2}^{\prime }}\rho _{s_{2}^{\prime
}t_{2}^{\prime },s_{1}^{\prime }t_{1}^{\prime }}^{(\pm )}(x,O(R^{-1})\mathbf{%
k}_{T},O(R^{-1})\mathbf{n})(R^{-1})_{t_{1}^{\prime }t_{1}}S_{s_{1}^{\prime
}s_{1}}(R^{-1})
\end{equation*}%
\begin{equation}
=\rho _{s_{2}t_{2},s_{1}t_{1}}^{(\pm )}(x,\mathbf{k}_{T},\mathbf{n})\,.
\label{rho-covariant}
\end{equation}%
We easily obtain Eq.~(\ref{psi-psi-MF}) from Eqs.~(\ref{ME-1}), (\ref%
{rho-def-Pi}).

\section{$P$-parity constraints}

\label{rho-structures-appendix} \setcounter{equation}{0}

In this appendix we show that the matrix function $\rho^{(\pm)}$ satisfying
constraints (\ref{rho-P-symmetry}), (\ref{Pi-rho-constraint}) can be
represented in the form (\ref{rho-A}).

We remind that the matrix $\rho^{(\pm)}$ acts on 4-component Dirac spinors
and 2-component isospinors. Due to the property (\ref{rho-def-Pi}) we
effectively deal with 2-component spinors so that there are $2\times2=4$
independent spin structures in which $\rho^{(\pm)}$ can be decomposed. For
example, we can choose

\begin{equation}
1,\gamma _{5},i(\mathbf{a}_{T}{\mbox{\boldmath$\gamma$}})
\label{rho-spin-structures}
\end{equation}%
where $\mathbf{a}_{T}$ is an arbitrary 3-vector obeying%
\begin{equation}
\left( \mathbf{a}_{T}\mathbf{n}\right) =0\,.
\end{equation}%
Note that all structures (\ref{rho-spin-structures}) commute with the
projector $\Pi (\mathbf{n})$ (\ref{Pi-n-def}). Taking%
\begin{equation}
\mathbf{a}_{T}=\mathbf{k}_{T}\,,\,\left[ \mathbf{n}\times \mathbf{k}_{T}%
\right] \,,
\end{equation}%
we obtain the following basis of spin matrices in which $\rho ^{(\pm )}$ can
be decomposed: 
\begin{equation}
\{B_{\alpha }\}=\left\{ 1,\,\gamma _{5},i\,(\mathbf{k}_{T}\mathbf{{%
\mbox{\boldmath$\gamma$}}}),\,i\mathbf{n}\cdot \left[ \mathbf{k}_{T}\times 
\mathbf{{\mbox{\boldmath$\gamma$}}}\right] \right\} \,.
\end{equation}%
In the isospin space we can use the following basis:%
\begin{equation}
\{C_{\beta }\}=\left\{ 1,(\mathbf{k}_{T}\mathbf{{\mbox{\boldmath$\tau$}}}),(%
\mathbf{n{\mbox{\boldmath$\tau$}}}),\mathbf{n}\cdot \left[ \mathbf{k}%
_{T}\times \mathbf{{\mbox{\boldmath$\tau$}}}\right] \right\} \,.
\end{equation}%
Now we can build%
\begin{equation}
4_{spin}\times 4_{isospin}=16
\end{equation}%
spin-isospin structures%
\begin{equation}
B_{\alpha }C_{\beta }\Pi (\mathbf{n})
\end{equation}%
which can be used as a basis for the decomposition of $\rho ^{(\pm )}$.

The matrix $\rho ^{(\pm )}$ should satisfy $P$-parity property (\ref%
{rho-P-symmetry}). Note that among $B_{\alpha }$ and $C_{\beta }$ only one
half of structures are $P$-even and others are $P$-odd:
\begin{equation}
\{B_{\alpha }\}=\left\{ 1,\,i\,(\mathbf{k}_{T}\mathbf{{\mbox{\boldmath$%
\gamma$}}})\right\} _{P\mathrm{-even}}\cup \left\{ \,\gamma _{5},\,i\,\left( 
\mathbf{n}\cdot \left[ \mathbf{k}_{T}\times \mathbf{{\mbox{\boldmath$\gamma$}%
}}\right] \right) \right\} _{P\mathrm{-odd}}\,,
\end{equation}%
\begin{equation}
\{C_{\beta }\}=\left\{ 1,\mathbf{n}\cdot \left[ \mathbf{k}_{T}\times \mathbf{%
{\mbox{\boldmath$\tau$}}}\right] \right\} _{P\mathrm{-even}}\cup \left\{ (%
\mathbf{k}_{T}\mathbf{{\mbox{\boldmath$\tau$}}}),(\mathbf{n{%
\mbox{\boldmath$\tau$}}})\right\} _{P\mathrm{-odd}}\,.
\end{equation}%
Therefore only 8 structures $B_{\alpha }C_{\beta }$ are $P$-even:%
\begin{equation}
\left\{ 1,\,i\,(\mathbf{k}_{T}\mathbf{{\mbox{\boldmath$\gamma$}}})\right\}
\otimes \left\{ 1,\mathbf{n}\cdot \left[ \mathbf{k}_{T}\times \mathbf{{%
\mbox{\boldmath$\tau$}}}\right] \right\} \,,  \label{P-even-str1}
\end{equation}%
\begin{equation}
\left\{ \,\gamma _{5},\,\,i\left( \mathbf{n}\cdot \left[ \mathbf{k}%
_{T}\times \mathbf{{\mbox{\boldmath$\gamma$}}}\right] \right) \right\}
\otimes \left\{ (\mathbf{k}_{T}\mathbf{{\mbox{\boldmath$\tau$}}}),(\mathbf{n{%
\mbox{\boldmath$\tau$}}})\right\} \,.  \label{P-even-str2}
\end{equation}%
Thus the matrix $\rho ^{(\pm )}$ obeying the relation (\ref{rho-P-symmetry})
can be decomposed in these structures (\ref{P-even-str1}), (\ref{P-even-str2}%
). However, for our purposes we prefer to replace one of these structures by
a linear combination. To this aim we use the following identity%
\begin{equation}
\left( \mathbf{n}\cdot \lbrack \mathbf{{\mbox{\boldmath$\gamma$}}}\times 
\mathbf{k}_{T}]\right) (\mathbf{k}_{T}\mathbf{{\mbox{\boldmath$\tau$}}}%
)+\left( \mathbf{k}_{T}\cdot \lbrack \mathbf{{\mbox{\boldmath$\tau$}}}\times 
\mathbf{n}]\right) (\mathbf{k}_{T}\mathbf{{\mbox{\boldmath$\gamma$}}}%
)=\left( \mathbf{{\mbox{\boldmath$\tau$}}}\cdot \lbrack \mathbf{n}\times 
\mathbf{{\mbox{\boldmath$\gamma$}}}]\right) |\mathbf{k}_{T}|^{2}\,.
\label{str-id}
\end{equation}%
Indeed, in the 3-dimensional space the antisymmetrization with respect to
four indices always gives zero:%
\begin{equation}
\varepsilon ^{\lbrack jkl}\delta ^{m]n}=0\,.
\end{equation}%
Contracting this equality with%
\begin{equation}
n^{j}\gamma ^{k}k_{T}^{l}\tau ^{m}k_{T}^{n}\,,
\end{equation}%
we obtain Eq.~(\ref{str-id}). Note that both terms entering in the LHS of
Eq. (\ref{str-id}) appear among the structures (\ref{P-even-str1}), (\ref%
{P-even-str2}). The identity (\ref{str-id}) means that in the basis (\ref%
{P-even-str1}), (\ref{P-even-str2}) we can replace%
\begin{equation}
\left( \mathbf{n}\cdot \left[ \mathbf{k}_{T}\times \mathbf{{%
\mbox{\boldmath$\tau$}}}\right] \right) (\mathbf{k}_{T}\mathbf{{%
\mbox{\boldmath$\gamma$}}})
\end{equation}%
by the other independent structure%
\begin{equation}
\left( \mathbf{{\mbox{\boldmath$\tau$}}}\cdot \lbrack \mathbf{n}\times 
\mathbf{{\mbox{\boldmath$\gamma$}}}]\right) \,.
\end{equation}%
With this change of the basis (\ref{P-even-str1}), (\ref{P-even-str2}) we
arrive at the decomposition (\ref{rho-A}) for $\rho ^{(\pm )}$. Note that
due to Eq. (\ref{V-r-scalar}) representation (\ref{rho-A}) for $\rho ^{(\pm
)}$ satisfies condition (\ref{rho-covariant}).

\section{Calculation of $R$ integrals}

\label{R-integrals-appendix} \setcounter{equation}{0}

In this appendix we derive Eq.~(\ref{int-R-res}).

We have according to Eq.~(\ref{rho-A})%
\begin{equation}
\rho _{s_{2}t_{2},s_{1}t_{1}}^{(\pm )}(x,\mathbf{k}_{T},\mathbf{n})=\delta
_{t_{2}t_{1}}\rho _{s_{2}s_{1}}^{(\pm )0}(x,\mathbf{k}_{T},\mathbf{n})+\tau
_{t_{2}t_{1}}^{a}\rho _{s_{2}s_{1}}^{(\pm )a}(x,\mathbf{k}_{T},\mathbf{n})\,.
\end{equation}%
where
\begin{equation}
\rho ^{(\pm )0}(x,\mathbf{k}_{T},\mathbf{n})=\left\{ V_{1}^{(\pm
)}+V_{5}^{(\pm )}i(\mathbf{k}_{T}\mathbf{{\mbox{\boldmath$\gamma$}}}%
)\right\} \Pi (\mathbf{n})\,,  \label{rho-0-V}
\end{equation}%
\begin{equation*}
\rho ^{(\pm )a}(x,\mathbf{k}_{T},\mathbf{n})=\left\{ V_{2}^{(\pm )}\gamma
_{5}k_{T}^{a}+V_{3}^{(\pm )}\gamma _{5}n^{a}+V_{4}^{(\pm )}\left[ \mathbf{%
n\times k}_{T}\right] ^{a}\right.
\end{equation*}%
\begin{equation}
\left. +V_{6}^{(\pm )}i\left( \mathbf{n}\cdot \lbrack \mathbf{{%
\mbox{\boldmath$\gamma$}}}\times \mathbf{k}_{T}]\right) n^{a}+V_{7}^{(\pm
)}i\left( \mathbf{n}\cdot \lbrack \mathbf{{\mbox{\boldmath$\gamma$}}}\times 
\mathbf{k}_{T}]\right) k_{T}^{a}+V_{8}^{(\pm )}i\left( \mathbf{n\times {%
\mbox{\boldmath$\gamma$}}}\right) ^{a}\right\} \Pi (\mathbf{n})\,.
\label{rho-a-V}
\end{equation}%
Then
\begin{equation*}
 R_{t_{2}t_{2}^{\prime }}\rho _{s_{2}t_{2}^{\prime
},s_{1}t_{1}^{\prime }}^{(\pm )}(x,\mathbf{k}_{T},\mathbf{n}%
)(R^{-1})_{t_{1}^{\prime }t_{1}}
\end{equation*}%
\begin{equation}
=\delta _{t_{2}t_{1}}\rho _{s_{2}s_{1}}^{(\pm )0}(x,\mathbf{k}_{T},\mathbf{n}%
)+(R\tau ^{a}R^{-1})_{t_{2}t_{1}}\rho _{s_{2}s_{1}}^{(\pm )a}(x,\mathbf{k}%
_{T},\mathbf{n})\,.
\end{equation}%
Here%
\begin{equation}
R\tau ^{a}R^{-1}=O^{ba}(R)\tau ^{b}\,.
\end{equation}%
and the matrix $O^{ba}(R)$ was defined in Eq.~(\ref{O-ab-def}). Now we have%
\begin{equation*}
 R_{t_{2}t_{2}^{\prime }}\rho _{s_{2}t_{2}^{\prime
},s_{1}t_{1}^{\prime }}^{(\pm )}(x,\mathbf{k}_{T},\mathbf{n}%
)(R^{-1})_{t_{1}^{\prime }t_{1}}
\end{equation*}%
\begin{equation}
=\delta _{t_{2}t_{1}}\rho _{s_{2}s_{1}}^{(\pm )0}(x,\mathbf{k}_{T},\mathbf{n}%
)+O^{ba}(R)(\tau ^{b})_{t_{2}t_{1}}\rho _{s_{2}s_{1}}^{(\pm )a}(x,\mathbf{k}%
_{T},\mathbf{n})\,.
\end{equation}%
For longitudinally polarized states of nucleon described by wave functions (%
\ref{phi-D-def}) with $I=S=1/2$, we find using equations of Appendix \ref%
{Wigner-functions-appendix}:%
\begin{equation}
\int dR\left[ \phi _{I_{3}^{\prime }S_{3}^{\prime }}(R)\right] ^{\ast }\phi
_{I_{3}S_{3}}(R)O^{ba}(R)=-\frac{1}{3}(\tau ^{b})_{I_{3}^{\prime
}I_{3}}(\tau ^{a})_{S_{3}^{\prime }S_{3}}\,.  \label{phi-phi-O}
\end{equation}%
In the case of an arbitrary polarization $S^{\mu }$ of the
nucleon in the rest frame
(\ref{S-rest}) we have the following generalization of Eq.~(\ref%
{phi-phi-O}):%
\begin{equation}
\int dR\phi _{I_{3},\mathbf{S}}^{\ast }(R)\phi _{I_{3}\mathbf{,S}%
}(R)O^{ba}(R)=-\frac{1}{3}(\tau ^{b})_{I_{3}I_{3}}S^{a}=-\frac{2}{3}\delta
_{b3}I_{3}S^{a}\,.
\label{int-3}
\end{equation}%
Now we can compute the integral in the LHS of Eq.~(\ref{rho-A})
using Eqs.~(\ref{phi-orthonormal}), (\ref{int-3}):
\begin{equation*}
\int dR\phi _{I_{3},\mathbf{S}}^{\ast }(R)\phi _{I_{3}\mathbf{,S}}(R)\left[
R_{t_{2}t_{2}^{\prime }}\rho _{s_{2}t_{2}^{\prime },s_{1}t_{1}^{\prime
}}^{(\pm )}(x,\mathbf{k}_{T},\mathbf{n})(R^{-1})_{t_{1}^{\prime }t_{1}}%
\right]
\end{equation*}%
\begin{equation*}
=\int dR\phi _{I_{3},\mathbf{S}}^{\ast }(R)\phi _{I_{3}\mathbf{,S}}
\end{equation*}%
\begin{equation*}
\times \left[ \delta _{t_{2}t_{1}}\rho _{s_{2}s_{1}}^{(\pm )0}(x,\mathbf{k}%
_{T},\mathbf{n})+O^{ba}(R)(\tau ^{b})_{t_{2}t_{1}}\rho _{s_{2}s_{1}}^{(\pm
)a}(x,\mathbf{k}_{T},\mathbf{n})\right]
\end{equation*}%
\begin{equation*}
=\delta _{t_{2}t_{1}}\rho _{s_{2}s_{1}}^{(\pm )0}(x,\mathbf{k}_{T},\mathbf{n}%
)-\frac{2}{3}\delta _{b3}I_{3}S^{a}(\tau ^{b})_{t_{2}t_{1}}\rho
_{s_{2}s_{1}}^{(\pm )a}(x,\mathbf{k}_{T},\mathbf{n})
\end{equation*}%
\begin{equation}
=\delta _{t_{2}t_{1}}\rho _{s_{2}s_{1}}^{(\pm )0}(x,\mathbf{k}_{T},\mathbf{n}%
)-\frac{2}{3}I_{3}(\tau ^{3})_{t_{2}t_{1}}S^{a}\rho _{s_{2}s_{1}}^{(\pm
)a}(x,\mathbf{k}_{T},\mathbf{n})\,.
\end{equation}

Taking into account Eqs.~(\ref{rho-0-V}), (\ref{rho-a-V}), we find%
\begin{equation*}
\int dR\phi _{I_{3},\mathbf{S}}^{\ast }(R)\phi _{I_{3}\mathbf{,S}}\left[
R_{t_{2}t_{2}^{\prime }}\rho _{s_{2}t_{2}^{\prime },s_{1}t_{1}^{\prime
}}^{(\pm )}(x,\mathbf{k}_{T},\mathbf{n})(R^{-1})_{t_{1}^{\prime }t_{1}}%
\right]
\end{equation*}%
\begin{equation*}
=\left( \delta _{t_{2}t_{1}}\left[ V_{1}^{(\pm )}+V_{5}^{(\pm )}i(\mathbf{k}%
_{T}\mathbf{{\mbox{\boldmath$\gamma$}}})\right] \right.
\end{equation*}%
\begin{equation*}
-\frac{2}{3}I_{3}(\tau ^{3})_{t_{2}t_{1}}S^{a}\left\{ V_{2}^{(\pm )}\gamma
_{5}k_{T}^{a}+V_{3}^{(\pm )}\gamma _{5}n^{a}\right.
\end{equation*}%
\begin{equation*}
+V_{4}^{(\pm )}\left[ \mathbf{n\times k}_{T}\right] ^{a}+V_{6}^{(\pm
)}i\left( \mathbf{n}\cdot \lbrack \mathbf{{\mbox{\boldmath$\gamma$}}}\times 
\mathbf{k}_{T}]\right) n^{a}
\end{equation*}%
\begin{equation}
\left. \left. +V_{7}^{(\pm )}i\left( \mathbf{n}\cdot \lbrack \mathbf{{%
\mbox{\boldmath$\gamma$}}}\times \mathbf{k}_{T}]\right)
k_{T}^{a}+V_{8}^{(\pm )}i\left( \mathbf{n\times {\mbox{\boldmath$\gamma$}}}%
\right) ^{a}\right\} \Pi (\mathbf{n})\right) _{s_{2}s_{1}}\,.
\label{int-R-res-0}
\end{equation}%
Now we can derive Eq.~(\ref{int-R-res}) from Eq.~(\ref{int-R-res-0}).

\end{document}